\documentclass[preprint,12pt]{elsarticle}

\usepackage{amssymb}
\usepackage{graphicx} 

\usepackage{float} 
\usepackage{amsmath}
\usepackage{mathtools}
\usepackage[margin=1in]{geometry}
\newcommand{\abs}[1]{\lvert#1\rvert}
\usepackage{lineno}

\journal{Extreme Mechanics Letters}

\begin{document}
	
	\begin{frontmatter}
		
		\title{Twist-Coupled Kirigami Cellular Metamaterials and Mechanisms}
		
		\author[label1]{Nigamaa Nayakanti}
		\address[label1]{Department of Mechanical Engineering, Masssachusetts Institute of Technology, Cambridge, MA,02139}
		
		\author[label1,label5]{Sameh H. Tawfick}
		\address[label5]{Department of Mechanical Science and Engineering, University of Illinois - Urbana Champaign, IL}
		\author[label1]{A. John Hart$^*$}
		
		
		
		
		\begin{abstract} 
			
			
	Manipulation of thin sheets by folding and cutting offers opportunity to engineer structures with novel mechanical properties, and to prescribe complex force-displacement relationships via material elasticity in combination with the trajectory imposed by the fold topology. We study the mechanics of cellular Kirigami that rotates upon compression, which we call Flexigami; the addition of diagonal cuts to an equivalent closed cell permits its reversible collapse without incurring significant tensile strains in its panels. Using finite-element modeling and experiment we show how the mechanics of flexigami is governed by the coupled rigidity of the panels and hinges and we design flexigami to achieve reversible force response ranging from smooth mono-stability to sharp bi-stability. We then demonstrate the use of flexigami to construct laminates with multi-stable behavior, a rotary-linear boom actuator, and self-deploying cells with activated hinges. Advanced digital fabrication methods can enable the practical use of flexigami and other metamaterials that share its underlying principles, for applications such as morphing structures, soft robotics and medical devices. 
		\end{abstract}
	\end{frontmatter}
	
	

\section*{Introduction}

Origami and Kirigami provide a means of transforming thin sheets into forms with unique mechanical properties, including extreme stretchability and multi-stability (\cite{Waitukaitis2014},\cite{Rafsanjani2015},\cite{Shan2015},\cite{Schenk2013}).  These materials and structures have promising applications in robotics (\cite{Felton2014}, \cite{Martinez2013}, \cite{Cheng14}, \cite{Na2015}), space structures (\cite{Viquerat2014},\cite{Schenk2013a},\cite{Schenk2014a}), soft actuators (\cite{Rus2015},\cite{Yang2015},\cite{Lazarus2015},\cite{Overvelde2015},\cite{Shim2012}), photovoltaics (\cite{Lamoureux2015}), and other domains.  Traditional origami assumes that creases behave as perfect hinges with zero rotational stiffness, and that the panels and connecting hinges are perfectly rigid and have zero thickness. Though these assumptions have proven useful to represent the motion of many folded systems, the flexibility of sheets enables the transformation of folded shapes that, from a mathematical point of view, are rigidly non-foldable.  For example, using the principle of virtual folds, Silverberg et.al. \cite{Silverberg2015} concluded that the flexibility of the panel faces give rise to the multi-stable trajectory of the square-twist Origami tessellation .  This is one of many examples, including the Miura-Ori, and Resch patterns, whereby folding of a polygonal arrangement can enable complex geometric and mechanical transformations (\cite{Lv2014}).  

In addition to folded sheets with repeating unit cells, there have been many studies of folded cylinders (\cite{Nojima2000},\cite{Tachi2013},\cite{Cheung2014},\cite{Cai},\cite{Tachi2009},\cite{Filipov2015}) and tubes, which may be used as deployable booms and lightweight structural members.  These include tessellations of identical triangular panels arranged on helical strips (\cite{Guest1994},\cite{Guest1994a}), as well as variations of Miura-Ori sheets wrapped into tubes and assembled into cellular structures with perpendicular load bearing capabilities (\cite{Filipov2015}). Another cylindrical topology, the Kresling pattern, has a series of parallel diagonal creases approximating the pattern that arises during twist-buckling of a thin-walled cylinder (\cite{Martinez2012},\cite{Hunt2005},\cite{Jianguo2015}). However, for some geometries compression of these seamless folded cylinders results in permanent structural failure due to formation of kinks and creases in the triangular panels due to tensile strains that develop in the initial stages of compression. 

We study the mechanics of cut-relieved folded cylindrical cells which we call flexigami. The addition of diagonal cuts imparts flat foldability of cylindrical topologies without incurring kinking and also mechanical behavior that spans from smooth mono-stability to sharp bi-stability. We show that the mechanics of flexigami is governed by the interplay between its rigid kinematics and the elasticity of panels and folds. Moreover, the cut-relieved design permits accurate and efficient modeling of the energetics using geometrical mechanics as well as finite-element modeling, contrasting previous studies which used simplified triangular truss-like structures to approximate the mechanics. We then demonstrate the use of flexigami cells to build multi-stable structures, compact rotary-linear mechanisms, and collapsible lightweight lattices.

\section*{Geometry and construction of flexigami cells}

\begin{figure}[htpb!]
	\centering
	\includegraphics[width = 1\columnwidth]{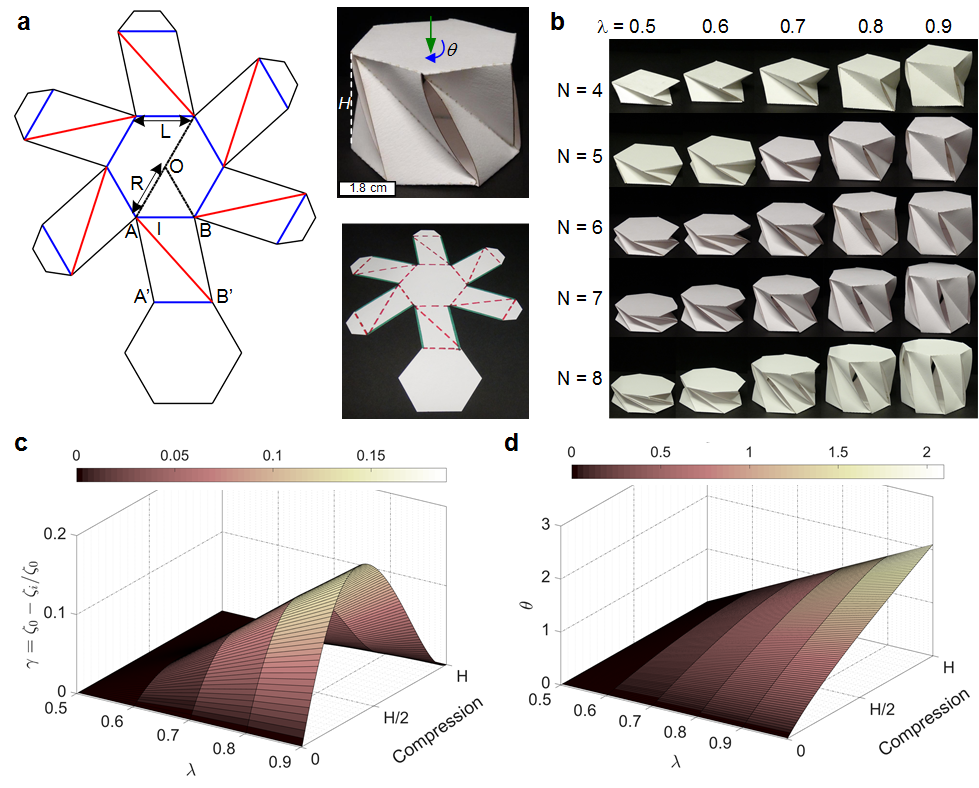}
	\caption{\textbf{Design of a twist-coupled kirigami (flexigami) cell and its kinematic analysis.} \textbf{a}: Unfolded unit cell pattern with edges marked in black, red lines indicating valley folds, and blue lines indicating mountain folds. \textbf{Insets}: Photographs of unfolded cell and folded paper cells with $N = 6,\ \lambda = 0.8$   \textbf{b}: Matrix of cells with varying $N, \lambda$. \textbf{c}: Variation in $\gamma$ as versus $\lambda$ and compression stroke. \textbf{d}: Relative rotation of cell versus $\lambda$ and compression stroke. }
	\label{Fig:Figure1}
\end{figure}

A flexigami cell has a N-sided polygonal base prism, surrounded by diagonally creased parallelograms (Figures~\ref{Fig:Figure1}a).  The cell geometry is defined by the geometric parameters of the regular polygon prescribed by such an arrangement: the number of sides of the polygon $\mathbf{N}$, the side length of the polygon $\mathbf{L}$, and a planar angular fraction $\mathbf{\lambda}$.  Each parallelogram is divided into a pair of triangles by a diagonal crease.  The creases are perforated, and then the planar pattern is folded and glued to form the enclosed cell.  When folded, the unit cell geometry is similar to a unit of the Kresling pattern, yet has cuts between each adjacent pair of triangles around the outer surface of its enclosed volume.  These cuts enable strain relief and reversible collapse of the cell, along with relative rotation of the top and bottom surfaces. 

In the folded yet fully open, stress-free configuration, three key geometric constraints can be imposed: (1) the N-sided faces remain planar and are only permitted to rotate about the vertical ($Z$) axis; (2) diagonal creases folded to valleys remain straight and their lengths are preserved; and (3) the free edges of the triangular panels ($AA'$, $BB'$) can assume any three-dimensional form, subject to the geometrical constraint of preserving their free length because their surfaces are considered to be developable.  As such, the circum-radius ($R$), diagonal crease length ($\xi$), and area of the triangle ($\zeta$) formed by joining the three points $AA'B'$ or $ABB'$ are given by

\begin{equation}
\begin{split}
& R =  \frac{L}{2\mathrm{cos}\left[\frac{\pi (N - 2)}{2 N}\right] } \\
& \\
& \xi = 2 R \ \mathrm{cos} \left[(1 - \lambda) \frac{\pi (N - 2)}{2 N}\right] \\
& \\
& \zeta  = 4 R^2 \ \mathrm{cos\left[\frac{\pi (N - 2)}{2 N}\right]}\  \mathrm{cos\left[(1 - \lambda)\frac{\pi (N - 2)}{2 N}\right]}\  \mathrm{sin\left[\lambda \frac{\pi (N - 2)}{2 N}\right]}
\end{split}
\end{equation}

Upon compression of the cell, the triangular panels must bend to accommodate the change in $\zeta$ while following the geometric constraints described above. The deviation ($\gamma$) of the panel geometry from a planar stress free configuration is expressed as

\begin{equation}
\begin{split}
& \gamma = \frac{\zeta_0 - \zeta_i}{\zeta_0}; 
\end{split}
\end{equation}

where, $\zeta_i$ : instantaneous area of the triangle $AA'B'$ or $ABB'$ and $\zeta_0$ : initial area of the triangle $AA'B'$ or $ABB'$.

Figure~\ref{Fig:Figure1}c depicts this variation for various values of $\lambda$, when $N = 7$,  $L = 30$. Here we observe that the change in $\gamma$ for all values of $\lambda$ is symmetric with respect to the relative height of a cell during compression. For $\lambda > 0.5$, we see that $\gamma$ exhibits a clear global maximum, which increases monotonically with $\lambda$.  Therefore, upon compression of the cell, the fractional change in $\zeta$ is accommodated by the out of plane bending of the panels.  Hence, for a given ($N, L$), the force required to compress the unit cell, and therefore its stiffness and peak force when bistable should increase with $\lambda$.  Because of the geometrical definition of the cell, its kinematic constraints are satisfied in exactly two configurations corresponding to completely open and closed states.  The existence of any intermediate configurations requires material deformations as well as opening and closing of the creases, suggesting that the mechanical response of the cell is governed by the panel and hinge stiffness.

\section*{Mechanical behavior of flexigami cells}

\begin{figure}[htpb!]
	\centering
	\includegraphics[width =  1\columnwidth]{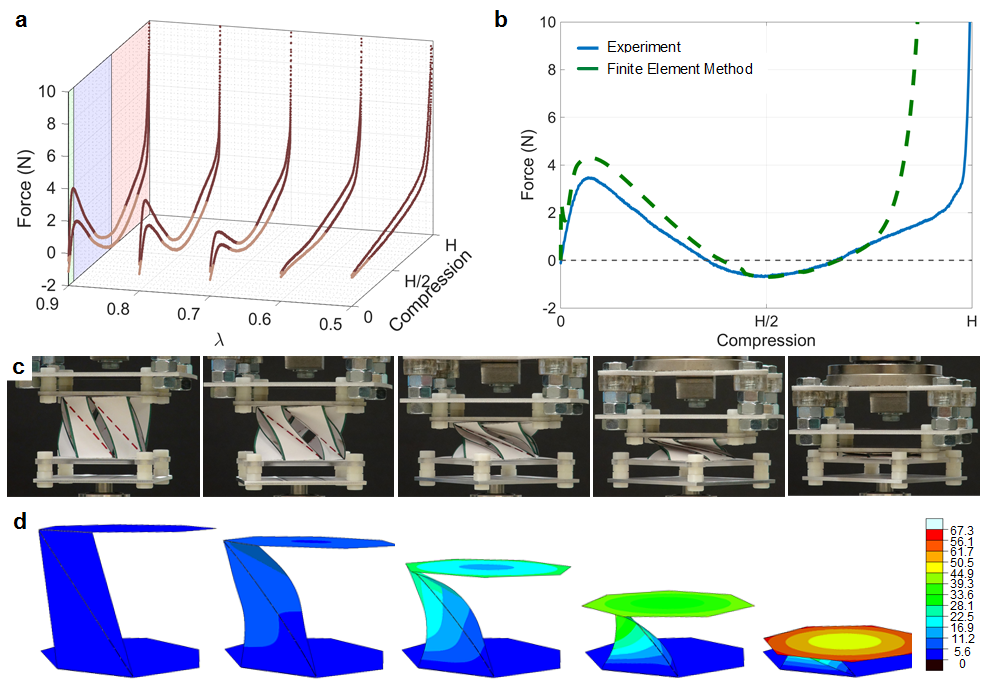}
	\caption{\textbf{Tension-compression testing and finite element modeling of flexigami cells.} \textbf{a}: Force-displacement reponse of paper cell ($N = 7,\ L = 30\text{mm}$) for $\lambda = 0.5, 0.6, 0.7, 0.8, 0.9$. \textbf{b}: Force-displacement response of corresponding finite element model, compared to experimental data. \textbf{c}: Photographs of paper cell during quasi-static compression, using fixture where bottom plate rotates freely and top plate rotation is fixed.  Green lines are marked along the edges of triangular panels highlight the bending deformations and red lines marking the creases remain straight. \textbf{d}: Corresponding frames from finite element simulation; color scale indicates relative displacement magnitude.}
	\label{Fig:Figure2}
\end{figure}

Displacement-controlled cyclic compression-tension tests (see \ref*{Sec:Methods}) on paper cells reveal the geometric tunability of mechanical behavior (Figure~\ref{Fig:Figure2}).  As shown in Figure~\ref{Fig:Figure2}a, for $\lambda = 0.5$ the cell opens and closes smoothly; upon compression, the force increases gradually past relative displacement of $H/2$ and stiffenss sharply only when the folded panels contact one another.  For $\lambda > 0.5$, the cell exhibits a snap-through behavior where it jumps from one stable equilibrium position to another.  In this case, the force first increases linearly until reaching a peak value; at the instant of the peak force ($F_P$), the cell snaps, taking a negative stiffness.  The force drops then to a local minimum value, and then the cell strengthens with continued compression due to monotonically increasing force required to fold the hinges and panel-panel contacts. 

For all the unit cells that exhibit snap-through behavior, their initial stiffness and peak force increase with $\lambda$ for a constant $N$ (Figure~\ref{Fig:Figure2}a). This agrees with the kinematic analysis showing that higher $\lambda$ results in more significant out-of-plane deformations of the panels ($\gamma$), while the hinges sweep through the same net fold angle (Figure~\ref{Fig:Figure1}c).  Similarly, for cells with $\lambda = \text{constant}$ (here 0.8), snap through behavior is always observed and the peak force increases with $N$ (Fig S2a).

When testing the flexigami cells, the top and bottom faces are held parallel, and only the bottom face is permitted to rotate. The diagonal cuts are essential to the smooth compression and bistability of flexigami cells. By contrast, seamless cells, become crumpled upon initial compression resulting in irreversible damage. This difference can be attributed to panel bending (Figure~\ref{Fig:Figure2}c, Video S1) and its influence on the bistability of cut-relieved cells. Without the cuts, the panels undergo combined stretching and bending; for instance, in Figure SI2 we shows the force-displacement response of a seamless cell (N = 7, $\lambda$ = 0.8). The closed cell exhibits higher initial peak force than the equivalent cut-relieved cell, but the sidewalls become wrinkled leading to collapse. In the subsequent cycles there is a loss of peak force and the force-displacement curves do not overlap as we would see in the case of cut-relieved cells. In tests of other geometries, we observe instability in the response after the peak force for higher values of $\lambda = 0.8$ \& $\lambda = 0.9$ where significant panel bending is otherwise common in cut-relieved cells. These observations suggest the influence of panel bending and irreversible nature of structural failure due to crumpling of the triangular panels.

A finite element model of a flexigami cell ($N$ = 7, $\lambda$ = 0.8) with a single panel is shown in Figure~\ref{Fig:Figure2}d (Video-S2).  Consistent with tensile tests on paper strips, the paper is modeled as an orthotropic material with $E_1$ = 6.8GPa and $E_2$= 3.1GPa (Fig. SI3).  Creases are modeled as torsional springs with constant stiffness.  The simulated response is compared with the experimental data for the same geometry in Figure~\ref{Fig:Figure2}b); it quite accurately predicts the initial stiffness, peak force, and negative force after snap-through, yet does not capture the stiffening upon panel-panel contact.  The finite element model also lets us visualize how, during compression, the strains in the creases increase monotonically, while the strains in the panels reach a maximum and subsequently decrease as the cell is compressed fully.  At the instant of maximum bending, the stresses developed in the panels increase in the direction away from the diagonal hinge. 

The paper cells also exhibit significant hysteresis in their compression-tension behavior.  Sequentially removing panels (i.e., single triangle pairs $AA'BB'$) from a cell predictably decreased the peak force, yet decreased the hysteresis even more-so (Fig.~SI2c).  Therefore, we conclude that friction between adjacent panels contributes significantly to the energy dissipation upon cyclic compression-tension .   Also, the mechanical behavior is insensitive to loading rate within bounds tested (25-200 mm/min), and unless otherwise noted tests were performed at 25mm/min (Fig.~SI2b).  

\section*{Interplay of panel and crease mechanics}
The interplay of panel bending and crease folding in twist-coupled Kirigami cells can be further understood using a geometric mechanics approach.  At each instantaneous height of the compressed cell, the total energy of the system ($E_T$), is the summation of the bending energy of panels ($E_B$) and energy stored in the creases ($E_C$).  The triangular panels are modeled as developable surfaces \cite{Developable2010}, and the governing pair of space curves are the free edge of the triangular panel ($\mathbf{R}_{AA'}(s)$) and the diagonal crease ($\mathbf{R}_{AB'}(s)$). In the deformed state, the diagonal crease remains straight ($\mathbf{R}_{AB'}(s)$), and the free edge of each triangular panel ($\mathbf{R}_{AA'}(s)$) is represented as

\begin{equation}
\mathbf{R}_{AA'}(s) = a_1 \mathrm{sin}(\pi s) + a_2 \mathrm{sin}(2\pi s) + a_3 \mathrm{sin}(3\pi s)
\end{equation}

Assuming isometric deformations of the triangular panels, the panels are thus developable surfaces.  The surface of each panel is then generated by joining the corresponding points on the two space curves $\mathbf{R}_{AA'}(s)$ and $\mathbf{R}_{AB'}(s)$ and is mathematically represented as

\begin{equation}
\mathbf{R}(s,v) = (1-v)\mathbf{R}_{AA'}(s) + v \mathbf{R}_{AB'}(s).
\label{Eq:RulingEq}
\end{equation}

The bending energy of a developable surface is proportional to the surface integral of the squared mean curvature (\cite{Dias2012},\cite{Love94}). Thus for each side of the cell,

\begin{equation}
E_B = 2 K_B \int \int H^2 d\mathrm{S}.
\label{Eq:BendingEnergy}
\end{equation}

\noindent Here, $K_B$ is the bending rigidity of the panels which is a function of Young's modulus ($E$), Poisson's ratio ($\nu$) and material thickness ($t$). 

\begin{figure}[htpb!]
	\centering
	\includegraphics[width = 1\columnwidth]{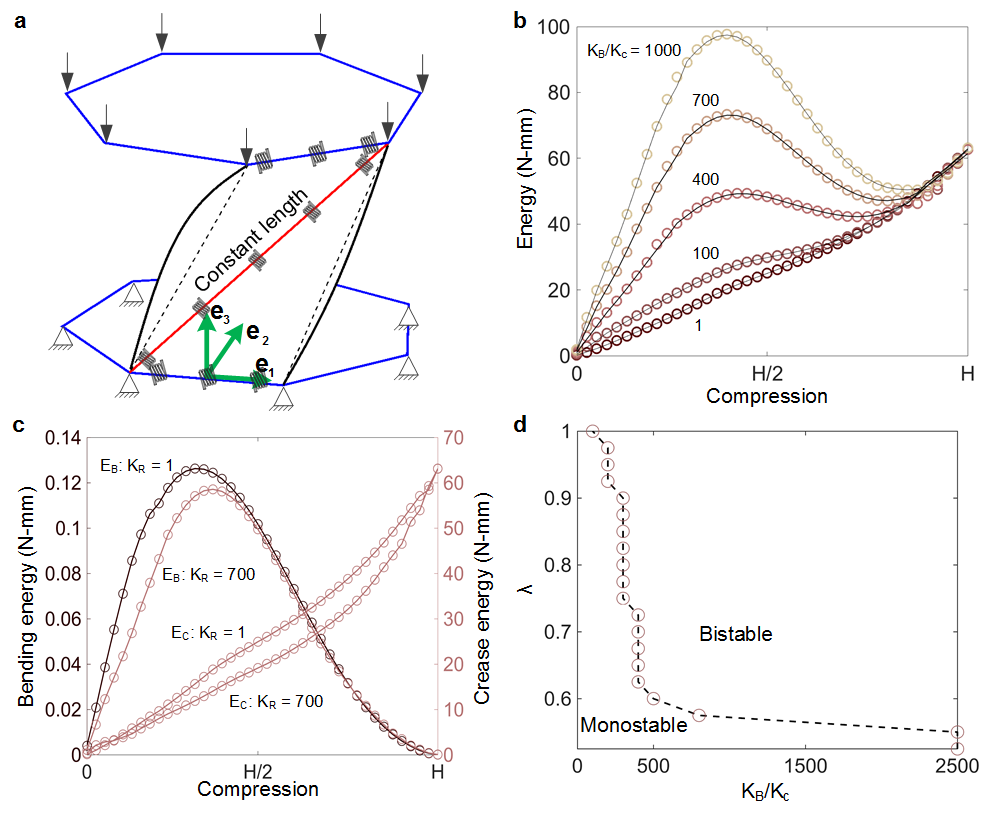}
	\caption{\textbf{Geometric mechanics analysis of flexigami unit cell.} \textbf{a}: Geometric model with constraints and displacement boundary conditions indicated; panels are flexible and creases are modeled as torsional springs. \textbf{b}: Predicted energetics of the cell as the ratio of panel to crease rigidity is varied, revealing a transition from monostability to bistability.  \textbf{c}: Contributions of panel (bending) and crease (folding) energies to the total energy for three values of $\mathrm{K_B/K_C}$. \textbf{d}: Phase diagram indicating the transition between monostable to bistable as a function of $\mathrm{K_B/K_C}$.}
	\label{Fig:Figure3}
\end{figure}

During compression, the instantaneous fold angle of the creases ($\theta_h$) is calculated and the energy stored in the creases is proportional to the square of the deviation in their angle from the rest position $(\theta_h\ - \theta_0)^2$. We therefore consider the creases to be linear elastic torsional springs (Figures~\ref{Fig:Figure3}a)(\cite{Lechenault2014}), whose spring stiffness is given by $K_C$.  Each side of the $N$ sided cell has three creases, two mountain folds and one diagonal valley fold. So, the crease energy associated with a single side is
\begin{equation}
E_C = \frac{K_{C}}{2} \sum_{i = 1}^{3}\left[\int \left[(\theta_{h_i}- \theta_{0_i})  \right] ^2 d\mathrm{l}\right],
\label{Eq:CreaseEnergy}
\end{equation}

where $\theta_{0_i}$ is the rest angle of the crease $i$, and $\theta_{h_i}$ is the instantaneous angle of the crease $i$.

By minimizing $E_T$ subject to prescribed kinematic and displacement boundary conditions,  we solve for the  equilibrium shape of the triangular panels at each instantaneous height during compression.  From this, we learn that for lower ratios of $K_R = K_B/K_C$, thus higher crease stiffnesses, the energy during compression increases monotonically (Figure~\ref{Fig:Figure3}b) implying mono-stability of the system. As $K_R$ increases further, an energy barrier develops, representing bi-stability.  By plotting the energy trajectories of panel bending and crease folding (Figure~\ref{Fig:Figure3}c), we see that the kinematic path of the cell causes the panels to bend through an energy maximum, while the crease energy increases smoothly throughout the compression stroke.  For each configuration ($N$, $\lambda$), we can therefore demarcate how $\lambda$ and $K_R$ determine whether the cell is monostable or bistable (Figure~\ref{Fig:Figure3}d).

Our geometric mechanics approach, wherein the panels are considered to be developable surfaces, is importantly different than prior efforts to explain the mechanics of either multi-stable origami (e.g., the square-twist, \cite{Silverberg2015}) using the principle of virtual folds or the equivalent seamless (e.g., Kresling) cylindrical shell using a simplified truss. In the virtual folds approach, the localized facet bending is approximated as a virtual fold and the panels are treated as rigid; in the truss approach, the fold pattern is simply analyzed as a network of linear elastic beams, and the panels are not considered. As a result, these approaches predict either a transition from mono-stability to bi-stability or obtain a condition on geometrical parameters for the structure to exhibit bi-stability as the fold pattern is varied. They donot accurately represent the scaling of energetics nor the coupling between crease folding and panel bending. Our approach involves direct definition of the load-free geometry of the structure along with the material properties of the panels and creases. Minimization of the total energy allows us to determine the equilibrium shapes of the panels and creases at the state of deformation, and therefore understand how the elasticity and kinematics together govern the structural response. Thus, this approach, along with appropriate formulation and computational solution methods, can be applied to any Origami or Kirigami structure.

\section*{Flexigami-based mechanisms and laminates}\label{Sec:Mechanisms}

\begin{figure}[htpb!]
	\centering
	\includegraphics[width = 1\columnwidth]{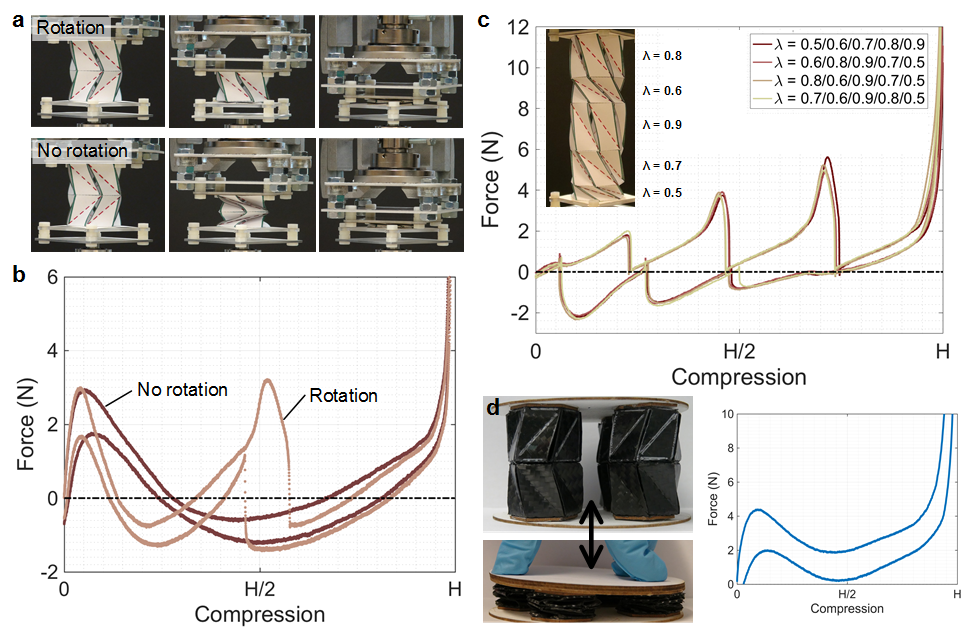}
	\caption{\textbf{Mechanics of stacked flexigami cells according to their configuration and twist constraint.} \textbf{a}: Photographs of cell pairs glued back-to-back, with and without relative rotation permitted upon compression.  \textbf{b}: Force-displacement curves corresponding to the images in (a) showing the same peak force in both cases, but simultaneous versus sequential snapping determined by the end constraint.  \textbf{c} Identical responses of four-cell stacks with arbitrary arrangements of cells with ($N =7$ and $\lambda = 0.6,\ 0.7,\ 0.8,\ 0.9$). \textbf{d} A small multi-layer array of cells fabricated from flexible carbon fiber sheets, with corresponding force-displacement curve of a monostable carbon fiber cell.}
	\label{Fig:Figure4}
\end{figure}

The above understanding of the geometric mechanics of flexigami cells enables the design of assemblies, laminates, and mechanisms having novel, nonlinear mechanical behavior and actuation response.  First, compression of two stacked cells of opposite chirality (Figure~\ref{Fig:Figure4}a), having at least one of its end surfaces constrained against rotation, results in sequential snapping combined with net rotation.  When both the top and bottom surfaces of the two-cell stack are constrained fully, the cells snap simultaneously in order to satisfy the constraint of zero net rotation. Both assemblies exhibit the same initial stiffness and peak force and can be compressed fully, yet the unconstrained assembly has three stable states and much greater hysteresis (Figure~\ref{Fig:Figure4}b).

It follows that single-chirality stacks of cells with identical $N$ and varied $\lambda$ have a number of stable states equal to the number of cells (Video-S3), yet their force-displacement curve is independent of the stacking order (Figure~\ref{Fig:Figure4}c).  Upon compression, the cells collapse in the sequential order of their peak force, thus a consecutively greater force is required to switch the stack to the next stable state.  Also, the negative force exerted by each snapping cell is compensated internally within the stack due to elasticity of the other cells, fully isolating the end constraints from the dynamics of snapping. 

It can therefore be appreciated that the engineerable mechanical response of twist-coupled flexigami cells, along with their large stroke, can enable the design of collapsible and deployable lightweight materials.  Practically, such will be limited by the formation of folded structures from structural materials, and in particular the construction of durable hinges that can reversibly endure large deformations. To demonstrate this possibility as well, we show an array of small unit cells made from a carbon fiber fabric (Carbitex).  This assembly (Figure~\ref{Fig:Figure4}d, Video S4; $N = 6,\ L = 15\mathrm{mm}, \ \lambda = 0.8$) is mono-stable as compared to a paper unit cell of identical geometry, as shown by the representative curve in Figure~\ref{Fig:Figure4}d.   The structure springs back upon reduction of the applied force because the crease energy overcomes the panel energy beyond the peak force, once again demonstrating the interplay of their relative contributions to the mechanical response.

\begin{figure}[htpb!]
	\vspace{-0.5cm}
	\centering
	\includegraphics[width = 1\columnwidth]{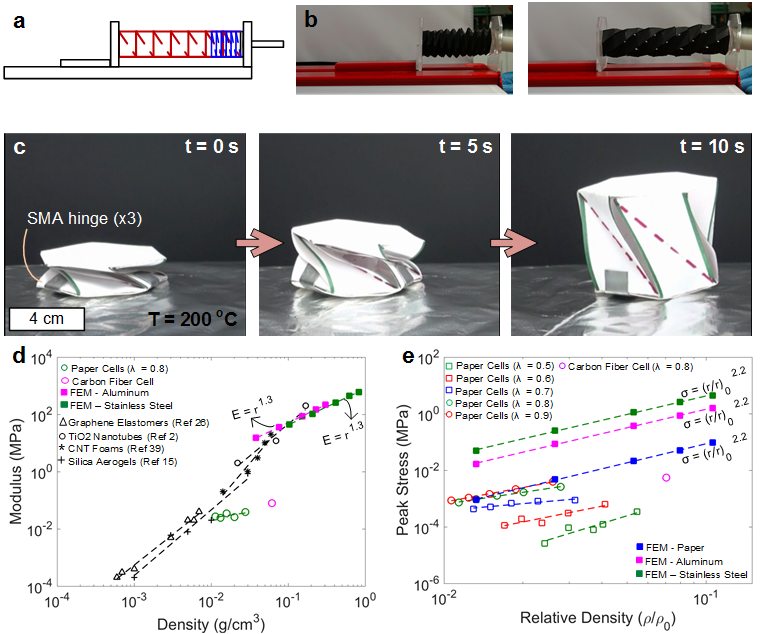}
	\caption{\textbf{Mechanisms built from twist-coupled flexigami cells, and scaling of mechanical behavior.} \textbf{a,b}: A linear axis motion system actuated by a stack of identical fleixgami cells constrained at both the ends. \textbf{c} A self-deploying cell using small ribbons of shape memory alloy attached to three of its bottom creases; the cell snaps open within 10 seconds after placement on a hot plate.  \textbf{d} Scaling of elastic modulus with mass density, comparing flexigami cells to low-density materials from literature, and extrapolating via FEM to potential fabrication of cells using structural metals.  \textbf{e} Relationships between relative density and peak stress at which the cell snaps to the collapsed state.}
	\label{Fig:Figure5}
\end{figure}

The coupling between linear extension and rotation of flexigami cells enables their configuration as linear-rotary actuators; for example, a stack of end-constrained cells us used to drive a simple linear stage (Figure~\ref{Fig:Figure5}, Video S5). Rotation of the base of the flexigami stack results in rapid deployment of the boom assembly, due to sequential opening of the internal cells.  In the specific case shown, we achieve a mean rotary to linear motion conversion of 10.8 mm/rad.  The maximum deployed length depends on ($N$,$L$,$\lambda$) and increases with each individually; variation in the ($\theta$-$H$) relation with $\lambda$ is shown in Figure~\ref{Fig:Figure1}d.  Contrasting cylindrical origami structures that offer continuous reversible deployment \cite{Filipov2015}, flexigami actuators, depending on their cell geometry and stacking, can be deployed either continuously or in multiple stages where each stage presents a stable equilibrium position. 

An alternative means of deploying flexigami cells is by direct actuation of the hinges, which is achieved herein using a shape memory alloy (SMA) foil (Nitinol, 0.127 mm thickness (product \# 045514), ThermoFisher Scientific).  Strips of SMA foil were first folded in half and trained to lay flat when heated past their transition temperature of 80$^\circ$C.  Three such strips were then glued to three alternate creases at the base of a six-sided cell, as shown in Figure~\ref{Fig:Figure5}c.  The cell, compressed in its stable closed position, was then placed on a hot plate pre-heated to  $200^\circ$C.  Within 10 seconds, the cell snaps to its stable extended position (Video S6), due to the force exerted by the foil hinges upon their phase transition.  This simple prototype shows the possibility to use twist-coupled cells in self-deploying structures, and to pursue further concepts for reversible actuation such as using antagonistic actuators at opposite ends of the cell, or using shape-memory hinges with two-way behavior.

The unique kinematics and properties of flexigami cells, including their extreme reversible strain, can also be considered in comparison to bulk cellular materials.  As such, we assess how the effective mechanical properties, namely the elastic modulus (Fig. ~\ref{Fig:Figure5}d) and peak stress (Fig. ~\ref{Fig:Figure5}e), scale with the relative density which is determined by the cell geometry, material density, and thickness.  The modulus and strength (i.e., the peak stress upon collapse) both scale with $\lambda$, and based on measurements of paper cells the properties are mutually maximized at the lowest $N$ (4) and highest $\lambda$ (0.9).  The design parameters ($\lambda$, $N$) govern the trajectory of panel bending and therefore scale the modulus which is related to the initial loading of the panels and the peak stress which is determined by the maximum bending deformation.  For the fabricated paper cells, the contribution of crease stiffness is negligible compared to panel bending; in the carbon fiber cells, similar scaling is observed but the crease stiffness is much larger leading to self-recovery after snap-through as shown earlier.  

Considered as a material, the modulus of flexigami cells exhibits near-linear ($E~\rho^{1.3}$) scaling with density, compared to conventional bend-dominated materials that exhibit quadratic scaling ($E~\rho^2$).  The modulus and relative density of paper cells are in fact comparable to those of previously studied ultralight materials, including silica aerogels (\cite{ADMA01103561},\cite{Worsley09},\cite{ADMA:ADMA201400249}) and graphene elastomers (\cite{Qiu12}), yet the sub-quadratic scaling is attractive for exploring the lower density regime.  Using a finite element model, we also predict the properties of cells with metal panels with identical geometry and thickness; for example, cells with aluminum and stainless steel panels are predicted to have modulus and strength 2 and 3  orders of magnitude greater than paper cells, respectively, at comparable density.  Fabrication of monolithic twist-coupled cellular materials using additive manufacturing techniques, both at microscopic and macroscopic scales (\cite{Bauer18022014}, \cite{Timothy14}) , also warrants future work.

\section*{Figures}

\newpage
\section*{Acknowledgements}\label{Sec:Ack}
Funding was provided by the National Science Foundation (EFRI-1240264) and by the U.S. Army Research Office through the Institute for Soldier Nanotechnologies under contract W911NF-13-D-0001.  We thank Matt Shlian for providing paper samples and initial models of a stack of collapsible folded cells, Sterling Watson for assistance in prototyping, Megan Roberts for previous work on fabrication and testing, and Sanha Kim, Abhinav Rao, and Justin Beroz for insightful discussions and advice.

\section*{Methods}\label{Sec:Methods}

The CAD drawings of the 2D cut pattern of individual flexigami unit cells were prepared using AutoCAD .  The patterns were cut from paper (Daler-Rowney canford; 150gm - 90lb, using a laser cutter (epilog mini 24), at 2\% power, 10\% frequency and 25\% speed of the laser cutter. To complete the unit cell, the pattern is folded sequentially along the creases, and the tabs on the triangular panels are glued using staples dot roller to one of the polygonal surfaces to obtain a closed flexigami unit cell. 

A Zwick tensile testing machine  with a 10kN load cell was used to carry out all the tensile tests being reported in this article. Figure. SI1 shows the experimental set-up. Here we see that the top platen is in series with the load cell. The bottom platen is fixed to a custom jig using c-clamps as shown. The custom made jig allows us to have rotation of the bottom surface of the unit cell with minimum inertia and has very high resistance against tilting because of any off centered loading on the rotating plate. This lets us record the force response of the unit cells accurately. Thumb screws located on the housing when tightened prevents the rotation of the plates and allow us to dynamically change the boundary conditions  applied to unit cells with one single set-up. 

\newpage
\section*{Supplementary Information} 
\renewcommand{\thefigure}{SI\arabic{figure}}
\setcounter{figure}{0}  
\begin{figure}[htpb!]
	\centering
	\includegraphics[width = 1\columnwidth]{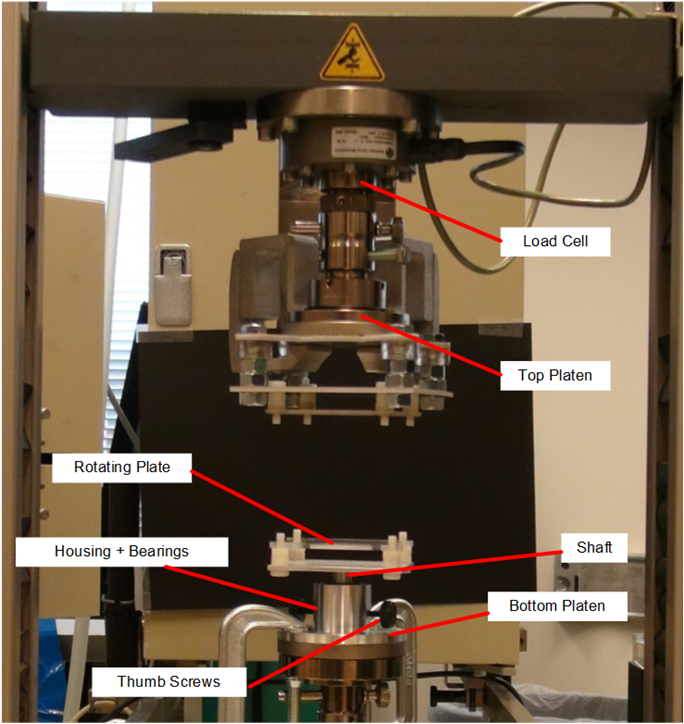}
	\caption{\textbf{Zwick mechanical testing machine used to perform all the tests reported in this article}Test set-up highlighting different components of the jigs used to which flexigami structures were attached. Inset shows the zoom-in view of the top and bottom plates to which flexigami structures are attached at four points.}
	\label{Fig:FigureS1}
\end{figure}

\subsection*{Compression of seamless Kresling cylinders}
\begin{figure}[htpb!]
	\centering
	\includegraphics[width = 1\columnwidth]{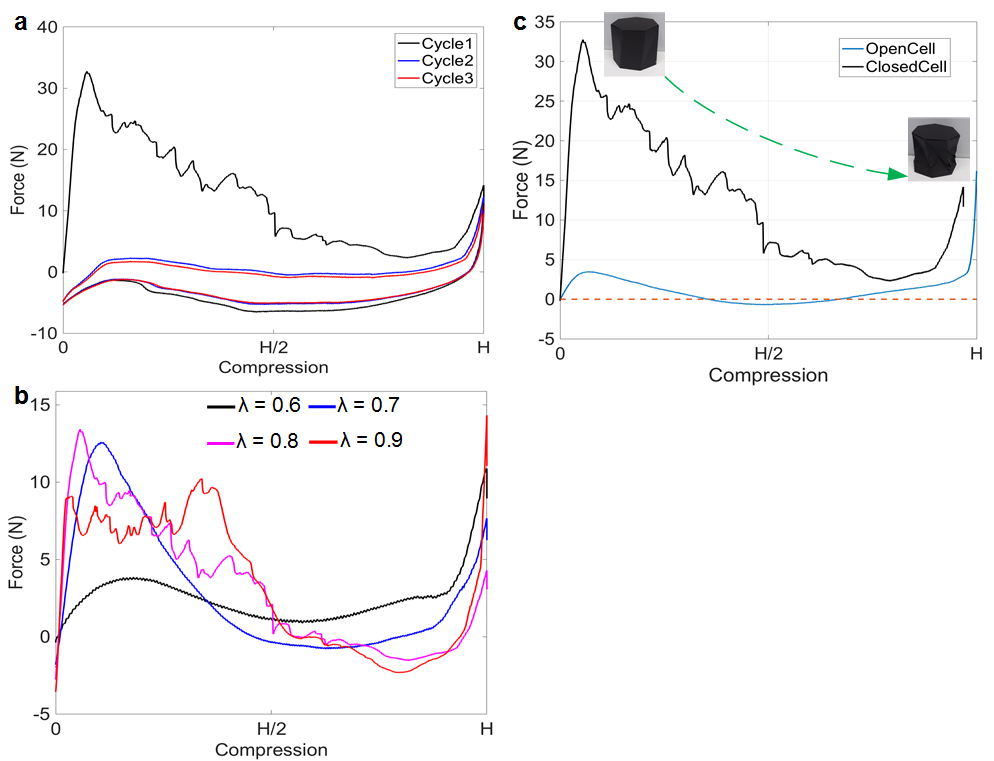}
	\caption{\textbf{Displacement controlled uni-axial compression-tension test on seamless cells. }\textbf{a:} Thee cycle test of a unit cell with N = 7 and $\lambda = 0.8$. \textbf{b:} Tensile force required to compress a unit cell with N = 7 and varying $\lambda$. \textbf{c:} Comparison between force-displacement response of open and seamless cells with N = 7 and $\lambda = 0.8$. Insets show the damage caused in seamless cylinders due to compression in a single cycle.}
	\label{Fig:FigureS4}
\end{figure}

Seamless cells with different geometric parameters (N and $\lambda$) are subjected to uni-axial compression and tension tests. In Figure~\ref{Fig:FigureS4} we show the force-displacement response of unit cells with N = 7 and $\lambda = 0.8$. Here we observe the response of the structure to compression in the first cycle is drastically different from its response to compression during subsequent cycles. Disturbances observed in the first compression cycle correspond to formation of creases or kinks on the triangular panels to relieve the strains developed in the process of compression. The peak force and stiffness of the structure drops substantially in subsequent cycles of compression and tension.

Unit cells of varying $\lambda$ for N = 7 are subjected to the same compression tension tests. Permanent deformation is less pronounced in the cells with lower $\lambda$ in comparison with higher values which is evident from the tests (Figure~\ref{Fig:FigureS4}b). This confirms that required deformation to accommodate the compression increases with increasing $\lambda$ and provision of cyclic cuts relieves the strains which would otherwise lead to formation of severe crumpling.

\subsection*{Compression of Flexigami Cells}

\subsubsection*{Effect of loading rate and panel contact}

\begin{figure}[htpb!]
	\centering
	\includegraphics[width = 1\columnwidth]{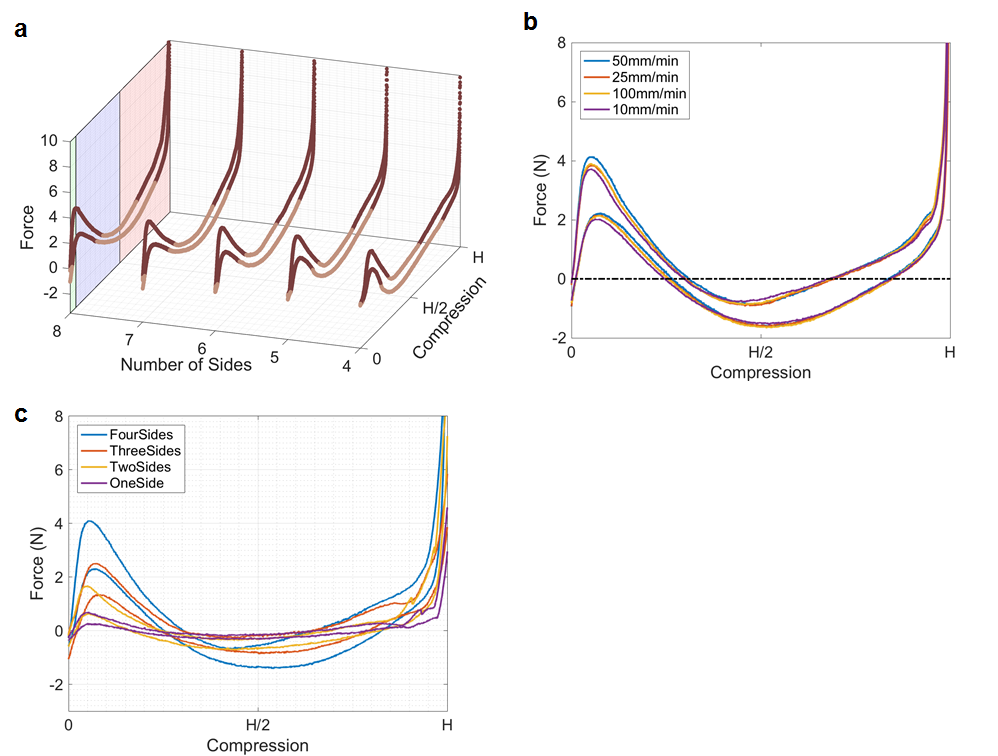}
	\caption{\textbf{Experimental data of displacement controlled uni-axial compressed tests on a unit cell}. \textbf{a: } Tensile force required to compressed and expand unit cells with $\lambda = 0.8 \& L = 30$mm.  \textbf{b: } Tensile force required to compress and expand a unit cell with $N = 4, L = 60\mathrm{mm},\ \lambda = 0.9$ under different test speeds. We see that when all the responses are overlaid, they form a very tight band and confirm that response of a unit cell is not significantly affected by the test speed. \textbf{c: } Force response of a unit cell with $N = 4, L = 60\mathrm{mm},\ \lambda = 0.9$ with 4,3,2,1 sides respectively. After each cycle of compression and tension, a side is removed from the unit cell and subjected to similar test conditions. Dwindling hysteresis and peak force is observed with reducing number of sides confirming the role of panel interactions and intrinsic hysteresis of the paper for being the reason for huge observed hysteresis.}
	\label{Fig:FigureS2}
\end{figure}

In order to confirm that the force response of the structures being tested is not affected by either the test speed or by the inertia of the rotating bottom plate, we carried out a series of tests on a unit cell with $N = 4, L = 60\mathrm{mm}, \lambda = 0.9$. Figure.~\ref{Fig:FigureS2} shows these results. Test speed is varied from $25\mathrm{mm/min}$ to $400\mathrm{mm}/min$. All the responses are overlapped on each other and form a very tight band (Figure.~\ref{Fig:FigureS2}b). This confirms that the effect of test speed and inertia of the bottom rotating plate on the response of a unit cell is negligible. All the force displacement results are reported at $25\mathrm{mm/min}$ test speed. 

Next to understand the role played by panel contact on the observed hysteresis, we again took a unit cell of $N = 4, L = 60\mathrm{mm}, \lambda = 0.9$. First with all the sides intact, we subjected it to displacement controlled uni-axial tension compression tests at varied speeds. After testing the specimen under different test speeds, we removed one of the sides and repeated the process. Next we removed the second side such that the remaining two sides are alternate with no interactions between them as the unit cell is being compressed. Next we brought it down to one side. All these tests are carried out under exactly same test parameters. Figure.~\ref{Fig:FigureS2}c shows the responses of these individual cases at 25mm/min.  Here we observe that as the number of sides reduces, peak force required to compress the unit cell and the amount of hysteresis in a cycle reduces significantly. Especially when  only one side is left, response of the unit cell in tension almost overlaps with the response in the compression part of the same cycle. So, we can affirmatively say that the large hysteresis observed in a unit cell is inherent to the structural components and the behavior of the paper itself but is not significantly affected by the hysteresis of testing machine.

\subsubsection*{Tensile properties of paper}

Papers are made from cellulose fiber and during the process of manufacturing, the axes of the fibers tend to be aligned parallel to the paper flow through the paper machine. This phenomenon leads to anisotropy in the mechanical properties of paper and  is generally considered to be orthotropic material. The thickness of the paper is much smaller compared to the other two in-plane directions. So, we can consider this as a case of plane stress. Under plane stress condition only the values of ($E_1$,$E_2$,$\nu_{12}$,$G_{12}$,$G_{13}$ and $G_{23}$) are required to define an orthotropic material. The Poisson's ratio $\nu_{21}$ is implicitly given as $\nu_{21} = (E_2/E_1)\nu_{12}$. The stress-strain relations for the in-plane components of the stress and strain are of the form

\begin{equation}
	\begin{Bmatrix}
		\varepsilon_1 \\
		\varepsilon_2 \\
		\gamma_{12}
	\end{Bmatrix} = \begin{bmatrix}
		1/E_1 & -\nu_{12}/E_1 & 0 \\
		-\nu_{12}/E_1 & 1/E_2 & 0 \\
		0 & 0 & 1/G_{12}
	\end{bmatrix} \begin{Bmatrix}
		\sigma_{11} \\
		\sigma_{22} \\
		\tau_{12}
	\end{Bmatrix}
\end{equation}

\begin{figure}[htpb!]
	\centering
	\includegraphics[width = 1\columnwidth]{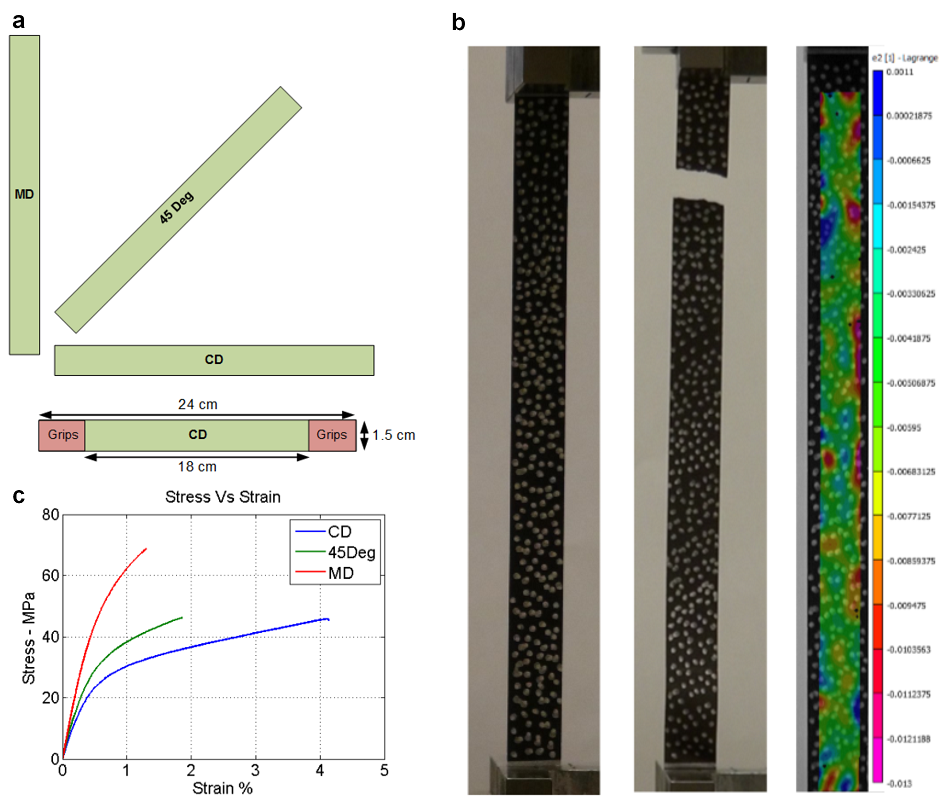}
	\caption{\textbf{a}: Representative figure of the directions along which the samples are cut from a Daler-Rowney Canford paper and the dimensions of the sample conforming to ISO standards for paper testing. \textbf{b}: Speckled sample before the test and after breakage along with representative set of axial  strain developed at the maximum extension. \textbf{c}: Stress-Strain response of the paper strips in the three measured directions. }
	\label{Fig:FigureS3}
\end{figure}

Specimens conforming to ISO standards were cut in three different orientations as shown in Figure.~\ref{Fig:FigureS3}a using a desktop mini Epilog laser cutter. 10 samples are cut in each direction with the specified dimensions and are speckled with silver sharpie to be able to use Digital Image Correlation (DIC) technique for the estimation of in-plane strains. 

In order to determine the in plane modulus and Poisson's ratio, we should have information about strains developed in the specimen in the longitudinal and transverse directions. Since, attaching a strain gauge would significantly affect the  properties of paper and would result in only one data point, we used the Digital Image Correlation (DIC) which is a non-contact optical strain measurement technique. 

Tests were conducted under displacement controlled conditions where the specimen was pulled at a constant velocity of 20mm/min with a 10KN load cell. A series of images of the specimen being deformed are captured and analyzed with Vic-2D. This provides us with the data of in-plane strains (Figure.~\ref{Fig:FigureS3}b). All the samples that break within 10mm of the clamping distance are rejected to meet with the ISO standards.

In-plane stress-strain curves of the Daler-Rowney canford (150gsm) paper are represented in Figure.~\ref{Fig:FigureS3}c . Table~\ref{Tab:MeanValues} lists the average Young's modulus along the two major directions (Machine direction and Cross direction) as well as shear modulus ($G_{12}$) and measured Poisson's ration $\nu_{12}$

\begin{table}[h!]
	\centering
	\begin{tabular}{|l|c|}
		\hline
		& Modulus (GPa)\\
		\hline
		& \\
		MD ($E_{1}$) & 6.83 \\
		\hline
		& \\
		CD ($E_{2}$) & 3.11 \\
		\hline 
		& \\
		$G_{12}$ & 2.17 \\
		\hline
		& \\
		$\nu_{12}$ & 0.23 \\
		\hline
	\end{tabular}
	\caption{Mean values of the modulus and Poisson's ratio}
	\label{Tab:MeanValues}
\end{table}

\subsection*{Finite-element simulations}

In the finite element model presented in the manuscript, the triangular panels are modeled as thin shell structures whose behavior is defined by its shell thickness, material density, and the in plane material properties as previously discussed. These properties also determine the individual bending rigidity ($K_b$) of the triangular panels. Both mountain and valley folds are modeled as linear elastic torsional springs whose behavior is determined by the torsional spring constant ($K_c$). Element type of STRI3 in Abaqus/Standard is used are used to mesh the triangular panels along with top and bottom polygonal surfaces. The element has three nodes, each with six degrees of freedom. The strains are based on thin plate theory, using small-strain approximation. Total of 13,960 STRI3 elements are used to mesh top and bottom plates; while, 6568 STRI3 elements are used to mesh each of the triangular panels. 
To model creases we use special purpose spring elements whose associated action is defined by the specified degrees of freedom involved. Type $\mathrm{I}$ and Type $\mathrm{II}$ creases are discretized into 9 and 17 points where these spring elements are acting. 

Specimen is then subjected to vertical displacement while fixing bottom plane (Figure.~\ref{Fig:FigureS1}) and imposing penalty to avoid node penetration between the two triangular panels which would come in contact in the process of compression. External work done on the entire specimen and reaction forces  at each and every node (from individual frames) on top and bottom panels of the specimen are obtained as output from the FE model. Vector summation of these reaction force components results in the force-displacement curve.

\subsection*{Geometric mechanics model}

Below, we detail the procedure of computing bending energy of panels which are considered to be developable as well as energy stored in the creases which are  modeled as torsional springs.

\subsubsection*{Bending Energy}
At any instantaneous height, we have two triangular panels and three creases corresponding to each of the $N$ sides. We need to parameterize the equations of these space curves to find the energy stored in each of them. Let the co-ordinates of each of the points be defined as follows

\begin{itemize}
	\item Point A: $\left[A_{x}, A_{y}, A_{z}\right]$; Point B: $\left[B_{x}, B_y, B_z\right]$;  
	\item Point A' : $\left[A'_x, A'_y, A'_z\right]$; Point B' : $\left[B'_x, B'_y, B'_z\right]$
\end{itemize}

Space curve $\mathbf{R}_{AA'}(s)$ is parameterized as: $\left[\abs{\overrightarrow{AA'}}s,\   \sum_{n = 1}^{3} a_n \mathrm{sin}(n \pi s), \ 0\right]$ in the co-ordinate system of ($\vec{e'_1}, \vec{e'_2}, \vec{e'_3}$) 

By co-ordinate transformation, representation of $\mathbf{R}_{AA'}(s)$ in the co-ordinate system of ($\vec{e_1}, \vec{e_2}, \vec{e_3}$) is :

\begin{equation}
	\begin{split}
		\mathbf{R}_{FE} & = \begin{bmatrix}
			Q_{11} & Q_{12} & Q_{13} \\
			Q_{21} & Q_{22} & Q_{23} \\
			Q_{31} & Q_{32} & Q_{33}
		\end{bmatrix}\begin{Bmatrix}
			\abs{\overrightarrow{AA'}}t \\
			\lambda sin(n \pi t) \\
			0
		\end{Bmatrix}  + \begin{bmatrix}
			A'_x \\
			A'_y \\
			A'_z
		\end{bmatrix}\\
		& \\
		& = \begin{bmatrix}
			Q_{11} \abs{\overrightarrow{AA'}} \ s+ Q_{12} \sum_{n = 1}^{3} a_n \mathrm{sin}(n \pi s) + A'_{x} \\
			Q_{21} \abs{\overrightarrow{AA'}} \ s+ Q_{22} \sum_{n = 1}^{3} a_n \mathrm{sin}(n \pi s)  + A'_{y}\\ 
			Q_{31} \abs{\overrightarrow{AA'}} \ s+ Q_{32} \sum_{n = 1}^{3} a_n \mathrm{sin}(n \pi s)  + A'_{z}
		\end{bmatrix}
	\end{split}
\end{equation}

where components of the transformation matrix $Q$ are given by : $$Q_{ij} = e_i \cdot e'_j$$

Similarly, space curve $\mathbf{R}_{AB'} (s)$ in the co-ordinate system ($\vec{e_1}, \vec{e_2}, \vec{e_3}$)  is given by

\begin{equation}
	\mathbf{R}_{AB'}(s) = \begin{bmatrix}
		B'_x + s (A_x - B'_x) \\
		B'_y + s (A_y - B'_y) \\
		B'_z + s (A_z - B'_z)		
	\end{bmatrix}
\end{equation}

Parametric equation of the triangular panel formed $AA'B'$ is given by

\begin{equation}
	\mathbf{R}_{AA'B'} (s,v) = (1 - v)\mathbf{R}_{AA'} (s)+ v \mathbf{R}_{AB'}(s)
\end{equation}

Similarly for the triangular panel formed by the three points $BB'A$, it is parametrically written as

\begin{equation}
	\mathbf{R}_{BB'A} (s,v) = (1-v)\mathbf{R}_{BB'} (s)+ v\mathbf{R}_{AB'}(s)
\end{equation}

Let $H_{AA'B'}$ and $H_{BB'A'}$ represent the mean curvature of the two surfaces, the total bending energy of them is:
\begin{equation}
	\begin{split}
		& \text{Bending Energy} = \\
		& K_B \int \int \left[H^2_{AA'B'} (s,v)   + H^2 _{BB'A'}(s,v)\right]\ \    \mathrm{d}s\  \mathrm{d}v
	\end{split}
\end{equation}

\subsubsection*{Crease Energy}	

For each of the $N$ sides, we have three creases: two creases correspond to mountain folds joining each of the triangular panels to either of the polygonal surfaces; and one Valley fold which is the diagonal crease joining the two triangular panels.  Creases are modeled as torsional springs with non-zero rest state which corresponds to zero crease energy. This non-zero rest state is the state in which a unit cell is in a fully open configuration. At any instantaneous height $h$, cosine of the angle  at any particular point on the crease is defined as dot product of the normals drawn on the two surfaces joining at that point. Following equations explain the procedure

\begin{itemize}
	\item Gradient of $\mathbf{R}_{AA'B'}$ with respect ot $v$ is 
	\begin{equation}
		\begin{split}
			& \left(\mathbf{R}_{AA'B'}\right)_v = -\mathbf{R}_{AA'}(s) + \mathbf{R}_{AB'}(s) \\
			& \\
			& \left(\mathbf{R}_{AA'B'}\right)_v (t,v)= \\
			& - \begin{bmatrix}
				Q_{11} \abs{\overrightarrow{AA'}} \ t+ Q_{12} \sum_{n = 1}^{3} a_n \mathrm{sin}(n \pi s)+ A'_x \\
				Q_{21} \abs{\overrightarrow{AA'}} \ t+ Q_{22}\sum_{n = 1}^{3} a_n \mathrm{sin}(n \pi s)   + A'_y\\ 
				Q_{31} \abs{\overrightarrow{AA'}} \ t+ Q_{32} \sum_{n = 1}^{3} a_n \mathrm{sin}(n \pi s)  + A'_z
			\end{bmatrix}  \\
			&+ \begin{bmatrix}
				B'_x + s (A_x - B'_x) \\
				B'_y + s (A_y - B'_y) \\
				B'_z + s (A_z - B'_z)		
			\end{bmatrix} 
		\end{split}
	\end{equation}
	\item Gradient of 	$\mathbf{R}_{AA'B'}$ with respect ot $s$ is 
	\begin{equation}
		\begin{split}
			&  \mathbf{R}_s (s,v) =  \\
			& (1-v)\begin{bmatrix}
				Q_{11} \abs{\overrightarrow{AA'}} + Q_{12} \sum_{n=1}^{3} na_n \pi\mathrm{cos}(n\pi s) \\
				Q_{21} \abs{\overrightarrow{AA'}} + Q_{22} \sum_{n=1}^{3} na_n \pi\mathrm{cos}(n\pi s) \\
				Q_{31} \abs{\overrightarrow{AA'}} + Q_{32} \sum_{n=1}^{3} na_n \pi\mathrm{cos}(n\pi s) 
			\end{bmatrix} \\
			& + v \begin{bmatrix}
				A_x -B'_x \\
				A_y - B'_y \\
				A_z - B'_z
			\end{bmatrix}	
		\end{split}
	\end{equation}
	
	\item Equation of normal ($\mathbf{N}$) to the tangent plane at a point ($s,v $) on the surface ($\mathbf{R}_{AA'B'}$) is given by
	\begin{equation}
		\mathbf{N}_{AA'B'} = \frac{\left(\mathbf{R}_{AA'B'}\right)_s \times \left(\mathbf{R}_{AA'B'}\right)_t}{\abs{\left(\mathbf{R}_{AA'B'}\right)_s \times \left(\mathbf{R}_{AA'B'}\right)_t}}
	\end{equation}		
	\item Similarly, equation of the normal ($\mathbf{N}$) to the tangent plane at a point  ($s,v$) on the surface $(\mathbf{R}_{BB'A})$ is: 
	\begin{equation}
		\mathbf{N}_{BB'A} = \frac{\left(\mathbf{R}_{BB'A}\right)_s \times \left(\mathbf{R}_{BB'A}\right)_t}{\abs{\left(\mathbf{R}_{BB'A}\right)_s \times \left(\mathbf{R}_{AA'B'}\right)_t}}
	\end{equation}
\end{itemize}

To compute crease energy of a crease, we need to track the change in the angle of the crease along its length and integrate it. Following equations give the instantaneous angle at each of the three creases and energy stored in them

\begin{itemize}
	\item	Energy stored in the crease $A'B'$ at height $h$
	\begin{equation}
		\begin{split}
			& \text{CreaseEnergy}_{A'B'} = \\
			& frac{K_C}{2} \int_{A'}^{B'} \left[(\theta_{A'B'})_h(0,v) - (\theta_{A'B'})_0(0,v)\right]^2 \mathrm{d}v
		\end{split}
	\end{equation}
	where $$(\theta_{A'B'})_h(0,v) = \mathrm{acos}\left((\mathrm{N}_{AA'B'} )_h(0,v)\cdot \vec{\mathbf{e}}_3\right)$$
	\item Energy stored in the crease $AB'$ at height $h$
	\begin{equation}
		\begin{split}
			& \text{Crease Energy}_{AB'} = \\
			& \frac{K_C}{2} \int_{A}^{B'} \left[(\theta_{AB'})_h(s,1) - (\theta_{AB'})_0(s,1)\right]^2 \mathrm{d}s
		\end{split}
	\end{equation}
	where $$(\theta_{AB'})_h(t,1) = \mathrm{acos}\left((\mathbf{N}_{AA'B'})_h(s,1) \cdot (\mathbf{N}_{BB'A})_h(s,1) \right)$$
	\item Energy stored int he crease $AB$ at height $h$
	\begin{equation}
		\begin{split}
			& \text{CreaseEnergy}_{AB} = \\
			& \frac{K_C}{2} \int_{A}^{B} \left[(\theta_{AB})_h(0,v) - (\theta_{AB})_0(0,v)\right]^2 \mathrm{d}v
		\end{split}
	\end{equation}
	where $$(\theta_{AB})_h(0,v) = \mathrm{acos}\left((\mathrm{N}_{BB'A} )_h(0,v)\cdot \vec{\mathbf{e}}_3\right)$$		
\end{itemize}

\begin{equation}
	\begin{split}
		& \text{Crease Energy} = \text{CreaseEnergy}_{A'B'} + \text{Crease Energy}_{AB'}   \\
		&+ \text{CreaseEnergy}_{AB}
	\end{split}
\end{equation}
So, for an $N$ sided unit cell, total energy of the system

\begin{equation}
	\begin{split}
		E_{Total} = & \ N\left[\text{Bending Energy} + \text{Crease Energy}\right] \\
		E_{Total} = &  N \int \int \left[H^2_{AA'B'} (s,v)   + H^2 _{BB'A'}(s,v)\right]\  \mathrm{d}s\  \mathrm{d}v  \\
		& + N \ \frac{K_C}{2} \int_{A'}^{B'} \left[(\theta_{A'B'})_h(0,v) - (\theta_{A'B'})_0(0,v)\right]^2 \mathrm{d}v \\
		& + N\  \frac{K_C}{2} \int_{A}^{B'} \left[(\theta_{AB'})_h(s,1) - (\theta_{AB'})_0st,1)\right]^2 \mathrm{d}s \\
		& + N\  \frac{K_C}{2} \int_{A}^{B} \left[(\theta_{AB})_h(0,v) - (\theta_{AB})_0(0,v)\right]^2 \mathrm{d}v
	\end{split}
\end{equation}

Minimization of $E_{Total}$ at each height $h$ provides us with the optimum values of ($a_i; i = 1,2,3$) which are then used to computed $\mathbf{R_{AA'}}$, $\mathbf{R}_{BB'}$, $\mathbf{R}_{AA'B'}$ and $\mathbf{R}_{BB'A}$ at that particular $h$.

\newpage
\bibliographystyle{model2-names.bst}
\bibliography{Literature2}


\end{document}